\documentclass[twocolumn,english]{IEEEtran}
\usepackage[T1]{fontenc}
\usepackage[latin9]{inputenc}
\usepackage{color}
\usepackage{array}
\usepackage{verbatim}
\usepackage{multirow}
\usepackage{amsmath,epsfig}
\usepackage{amssymb}
\usepackage{graphicx}
\usepackage{esint}

\ifCLASSINFOpdf
\else
  \DeclareGraphicsExtensions{.eps}
\fi
\usepackage{url}
\usepackage{color}
\usepackage[normalem]{ulem}

\begin{document}

%
\title{Cloud Instance Management and Resource Prediction For Computation-as-a-Service Platforms }
%
%
%

\author{Joseph Doyle,
        Vasileios Giotsas,
        Mohammad Ashraful Anam and~Yiannis~Andreopoulos,~\IEEEmembership{Senior~Member,~IEEE}
\thanks{This paper appears in the \textit{Proc. IEEE Int. Conf. on Cloud Engineering} (IC2E), Apr. 4-8, 2016, Berlin, Germany. This work was funded in part by  Innovate UK, project ACAME, grant no. 131983. The authors are with Dithen Ltd., 843 Finchley Road, London, NW11 8NA, U.K., http://www.dithen.com; email: \textbraceleft j.doyle, v.giotsas, russell.anam, i.andreopoulos\textbraceright@dithen.com.\ JD, MAA and YA are also with the Electronic and Electrical Engineering Department, University College London, Roberts Building, Torrington Place, London, WC1E 7JE, UK.}
\thanks{}}

%
%

\markboth{}%
{Shell \MakeLowercase{\textit{et al.}}: Bare Demo of IEEEtran.cls for Journals}
%



\maketitle

\begin{abstract}
Computation-as-a-Service (CaaS) offerings have gained traction in the last few years due to their effectiveness in balancing between the scalability of Software-as-a-Service and the customisation possibilities of Infrastructure-as-a-Service platforms.   To function effectively, a CaaS platform must have three key properties: \textit{(i)}\ reactive assignment of individual processing tasks to available cloud instances (compute units) according to availability and predetermined time-to-completion (TTC)\ constraints; (\textit{ii})\ accurate resource prediction; \emph{(iii)}  efficient control of the number of cloud instances servicing workloads, in order to optimize between completing workloads in a timely fashion and reducing resource utilization costs. In this paper, we propose three approaches that satisfy these properties (respectively): \textit{(i)} a service rate allocation mechanism based on proportional fairness and TTC constraints; \textit{(ii)} Kalman-filter estimates for resource prediction; and \textit{(iii)}\ the use of additive increase multiplicative decrease (AIMD) algorithms (famous for being the resource management in the transport control protocol) for the control of the number of compute units servicing workloads. The integration of our three proposals\ into a single CaaS platform is shown to provide for  more than 27\% reduction in Amazon EC2 spot instance cost against methods based on reactive resource prediction and  38\% to 60\%\  reduction of the billing cost  against the current state-of-the-art in CaaS platforms (Amazon Lambda and Autoscale).   \end{abstract}


%
\IEEEpeerreviewmaketitle

\section{Introduction}
%
%
%
%
\IEEEPARstart{C}{loud} Infrastructure-as-a-Service (IaaS) providers, such as Amazon Elastic Compute Cloud (EC2),  Google Compute Engine (GCE), VMware and Rackspace, now provide cloud instances, a.k.a. \textit{compute units} (CUs), i.e., pre-established sets of processor cores, memory, storage and operating systems, with yearly, daily, hourly or even minute-by-minute billing \cite{song2012optimal,zhang2014dynamic}. This has resulted in a variety of new  \textit{Platform-as-a-Service} (PaaS) \   and \textit{Software-as-a-Service} (SaaS)  offerings \cite{nan2012optimal,hobfeld2012challenges}. In a PaaS systems, a distributed computing environment (e.g., Apache Hadoop or mesos, Google App Engine, Microsoft Azure, etc.) is used to execute tasks on IaaS providers, albeit at the cost of porting the processing software to code that can be scaled-up by the PaaS infrastructure (for example converting the operations  to  a series of \texttt{Map()} and \texttt{Reduce()} steps in Hadoop).  In SaaS,  a specific set of
applications are licensed to customers (e.g., pre-established word processing software, a fixed set of video transcoding or video streaming toolboxes, etc.) either as a service on demand, through a
subscription, or in a pay-as-you-go model \cite{islam2012giving}.

\subsection{From Platform and Software-as-a-Service to Computation-as-a-Service}

This evolution of IaaS, PaaS and SaaS is now beginning to lead to \textit{Computation-as-a-Service} (CaaS) \cite{masiyev2012cloud},  where users can upload data (e.g., image, audio or video) files \textit{and} scripts or binary files, which can be executed by the CUs in the cloud directly, i.e., without the users having to set up and manage any infrastructure or convert their software to a format amenable to distributed computing environments. CaaS provides a useful compromise between the generality of IaaS and PaaS offerings and the  ease-of-use of SaaS: the end user can deploy and scale-up \textit{any} desktop data processing application of their choosing \textit{without} adapting its codebase. This differs from the case of SaaS,  in that the user can simply execute any Matlab, C/C++, Java, OpenCV,  Javascript/Python\ based code and scripts of their local platform on the CaaS\ platform without any modification by following a  simple set of rules. The CaaS platform can then handle the scheduling and parallelization of multiple workloads without any user intervention via the appropriate reservation (or bidding)\ of resources from IaaS providers, e.g., Amazon EC2 spot instance bidding or GCE\ CU reservation.

\subsection{ Related Work}
In order to provide for a viable service, a CaaS provider must be able to minimize the monetary cost incurred by the use of cloud CUs and schedule workloads to be executed in the most effective manner. To this end, there have been numerous recent proposals   for cloud resource management. Gandhi \textit{et. al.} propose their own version of Autoscale, which terminates servers that have been idle for more than a specified time, while consolidating jobs on less CUs to lower cost \cite{Gandhi2012autoscale}. Paya \textit{et. al.}  propose a system which expands on this  by using multiple sleep states to improve performance \cite{Paya2015loadbalancing}. Song \textit{et. al.} propose optimal allocation of CUs according to pricing and demand distributions \cite{song2012optimal}. Ranjan \textit{et. al.} investigate architectural elements of content-delivery networks with cloud-computing support \cite{ranjan2013mediawise}. Finally, Jung \textit{et. al.} propose using genetic algorithms for multi-user workload scheduling on various CUs \cite{jung2014estimation}.

Beyond resource allocation and scheduling, a major challenge in CaaS frameworks is the varying delay in the completion of various data processing workloads \cite{kontorinis2009statistical,andreopoulos2007adaptive,hobfeld2012challenges,gulisano2012streamcloud}. The processing delay primarily depends on: the specifics of the workload, the CU reservation mechanism employed, and the transport-layer jitter (if data is continuously transported to/from users and cloud providers) \cite{islam2012giving}. This is the primary reason why all real-world CaaS platforms only provide ``best effort" service level agreements (SLAs) for large workload execution without considering a predetermined time-to-completion (TTC)\ estimate. Recent research work on this front proposes the  derivation of viable schedules using particle swarm optimisation  \cite{Rodriguez2014deadlinescheduling} and the utilisation of the earliest-deadline first algorithm \cite{Mao2011deadlinescheduling}. While all such proposals are effective in their resource provisioning for TTC-abiding execution, they assume that the computation required to complete each workload can be accurately predicted by the system. However, this is unlikely to be the case in practice, particularly at the start of a workload's execution. Therefore, our proposal considers the realistic scenario where no estimates for the computational requirements are available at the beginning of each workload's execution; i.e., in conjunction with resource management, our framework performs an adaptive resource prediction \textit{during} the execution of each workload.

Finally, the first commercial CaaS\ offerings are now  beginning to emerge. The key representatives are: \textit{(i)}\ the recently-announced AWS Lambda service, where users can submit individual Javascript items and be billed at a fixed rate per 100ms of Lambda service usage under a best-effort SLA;  \textit{(ii)}\ PiCloud,\ a service for flexible scheduling of batch processing tasks via a terminal command line interface; \textit{(iii)}\ Parse, a software development environment for Javascript execution on cloud-computing infrastructures; and \textit{(iv)} Amazon EC2 Autoscale, a service that automatically scales application deployment over Amazon EC2 according to processor and network utilization constraints. In all these deployments, the comparative metric for workload analysis is the required processing time in terms of the number of seconds a single core was occupied until the workload is successfully completed. We therefore quantify the resource reservation in the IaaS via \textit{compute-unit seconds} (CUSs), i.e., the product of the total cores used  with the time they were reserved for, since charges will be applied for them from the infrastructure provider regardless of whether the CaaS system actually used them to their full capacity or not.

\subsection{ Contribution}

While the current research and commercial efforts in CaaS frameworks are a promising start, they do not consider  approaches for reactive prediction of the required CUSs to process submitted workloads, or assume that the CUS metric per workload is known \cite{hobfeld2012challenges,Rodriguez2014deadlinescheduling,Mao2011deadlinescheduling}. In addition, current CaaS\ frameworks do not consider on-demand CU provisioning (e.g., EC2 spot instances or GCE\ CUs with minute-level increments) under TTC\ constraints, where it is important to control \textit{both} the allocation and termination of new instances in order to reduce the infrastructure cost while providing for TTC-abiding execution. \ 

In this light, we present a CaaS platform that scales small and medium-level execution of data processing workloads to big data under TTC constraints. For example, algorithms for video transcoding, image classification, object recognition, etc., that run on small amounts of input images/videos on a desktop computing system can be directly scaled-up via the proposed platform (i.e., without any code modifications)  to operate on big datasets comprising millions of input images and videos, with \textit{a-priori} established completion times. Our CaaS platform meets the requirements for such large-scale data processing by  combining the following novel aspects:\  

\begin{enumerate}

\item
Each submitted workload is separated into individually-executable tasks, which are then allocated to available CUs with proportionally-fair scheduling in order to: \textit{(i)} maximize the available CU\ utilization and \textit{(ii)} abide by the confirmed TTC value for the workload.
The fine-grain partitioning of each workload into tasks allows for each user to check that the output results are being produced correctly by the platform during execution and cancel the workload execution if otherwise.

\item
The required CUSs until the completion of each task type in each workload are predicted based on Kalman-filter estimators, which are shown to significantly outperform other \textit{ad-hoc} estimators.  

\item 
Based on the proposed CUS prediction, we propose the use of the Additive Increase Multiplicative Decrease (AIMD)\ algorithm \cite{shorten2006positive}    for the allocation or termination of CUs according to the expected workload.
While AIMD is a well-known control mechanism for network resource utilization, e.g., within the transport control protocol (TCP), to the best of our knowledge, this is the first time it is proposed for CaaS provisioning.

\end{enumerate}\

 Finally, beyond describing our platform, we also provide free access to it\footnote{free usage with limits on the data volume and number of workload executions per user} at http://www.dithen.com.

The remainder of this paper is organized as follows.   Section \ref{Sec2} presents an overview of our platform, with Table \ref{tab:Notation-table} summarizing the utilized nomenclature. Sections \ref{Sec3} and \ref{Sec4}  present the key elements of the proposed CUS prediction  and AIMD framework, while Section \ref{Sec5} presents experimental results and comparisons of different CU\ allocation strategies for Amazon EC2 spot instances. Finally, Section \ref{Sec6}\ presents some concluding remarks.

\section{CaaS Platform Overview}\label{Sec2}

\begin{table}
\noindent \centering{}\protect\caption{\label{tab:Notation-table}Nomenclature and Notational Conventions.}
\begin{tabular}[t]{>{\centering}p{0.2\columnwidth}>{\raggedright}p{0.7\columnwidth}}
\hline 
\noalign{\vskip\doublerulesep}
\textbf{Key Concept} & \centering{}\textbf{Definition}\tabularnewline[\doublerulesep]
\hline 
\noalign{\vskip\doublerulesep}
\centering{}$t$ & monitoring time instant in the CaaS platform \tabularnewline[\doublerulesep]
\noalign{\vskip\doublerulesep}
\centering{}$W[t]$ & total workloads in Dithen at time instant $t$\tabularnewline[\doublerulesep]
\noalign{\vskip\doublerulesep}
\centering{}$M[t]$ & total data types at time \(t\)\tabularnewline[\doublerulesep]
\noalign{\vskip\doublerulesep}
\centering{}$m_{w,k}[t]$ & remaining data items of type $k$ to be processed within workload $w$ at time \(t\)  \tabularnewline[\doublerulesep]
\noalign{\vskip\doublerulesep}
\centering{}$I$ & total \textit{types} of instances in the cloud infrastructure\tabularnewline[\doublerulesep]
\noalign{\vskip\doublerulesep}
\centering{} $p_i$\   & compute units (CUs), i.e., processor cores, available within instance type \(i,1\leq i \leq I\) \tabularnewline[\doublerulesep]
\noalign{\vskip\doublerulesep}
\centering{}$n_i[t],N_\text{tot}$ & \textit{number} of instances of type \(i\) (\(1\leq i \leq I\)) reserved at time \(t\), total \textit{number} of CUs in Dithen \tabularnewline[\doublerulesep]
\noalign{\vskip\doublerulesep}
\centering{}$a_{i,j}[t]$\   & remaining time for the \(j\)th instance of type \(i\) before additional billing is incurred by the cloud provider  \tabularnewline[\doublerulesep]
\noalign{\vskip\doublerulesep}
\centering{}$c_{\text{tot}}[t],c_{\text{min}},c_{\text{max}}$\   & total compute-unit-seconds (CUSs) available in Dithen, and lower/upper limits for CUSs in Dithen  \tabularnewline[\doublerulesep]
\noalign{\vskip\doublerulesep}
$d_{w}[t]$ & time-to-completion (TTC) for workload \(w\) at time \(t\)\tabularnewline[\doublerulesep]
\noalign{\vskip\doublerulesep}
$\hat{b}_{w,k}[t]$ & prediction for the required CUSs to process a data item of type \(k\) of workload \(w\)\ \tabularnewline[\doublerulesep]
\noalign{\vskip\doublerulesep}
$r_{w}[t]$ & required CUSs for the completion of workload \(w\) \tabularnewline[\doublerulesep]
\noalign{\vskip\doublerulesep}
$s_{w}[t]$ & service rate, i.e., CUs allocated for workload \(w\) \tabularnewline[\doublerulesep]
\noalign{\vskip\doublerulesep}
$z_{w,k}[t],v_{w,k}[t]$ & CUS process and measurement noise instantiations of data type \(k\) of workload \(w\) \tabularnewline[\doublerulesep]
\noalign{\vskip\doublerulesep}
$\alpha,\beta$ & additive increase and multiplicative decrease parameters of AIMD \tabularnewline[\doublerulesep]
\hline 
\noalign{\vskip\doublerulesep}
\textbf{Notation} & \centering{}\textbf{Explanation}\tabularnewline[\doublerulesep]
\noalign{\vskip\doublerulesep}
\hline 
\noalign{\vskip\doublerulesep}
uppercase Roman letters & random variables\tabularnewline[\doublerulesep]
\noalign{\vskip\doublerulesep}
lowercase Greek letters & moments of probability distributions, stochastic parameters of Kalman filters, or AIMD and ARMA\ parameters\tabularnewline[\doublerulesep]
\noalign{\vskip\doublerulesep}
$\tilde b$ & measurement of quantity \(b\)\tabularnewline[\doublerulesep]
\noalign{\vskip\doublerulesep}
$\hat b$ & prediction estimate of quantity \(b\)\tabularnewline[\doublerulesep]
\noalign{\vskip\doublerulesep}
\hline 
\noalign{\vskip\doublerulesep}
\end{tabular}
\end{table}

\label{sec:ME}
Consider a CaaS platform where, for every monitoring instant $t$, $W[t]$ workloads have been submitted by its users and remain to be processed, with each workload  $w$ ($1\leq w \leq W[t]$) comprising $M[t]$ data types (e.g., images, videos, feature vectors, log files, etc.).\ In order to function in an efficient manner, per workload $w$ and data type $k$ within the workload, the CaaS platform keeps track of: the number of remaining elements to be processed, $m_{w,k}[t]$,
as well as the prediction estimates for the required CUSs to complete the processing,  $\hat{b}_{w,k}[t]$. Typically, the SLA for each workload includes execution within a predetermined TTC value,  $d_w[t]$, which is confirmed after an initial CUS prediction is available for the workload. To this end, the platform continuously keeps track of the required CUSs to complete each workload \(w\), $r_w[t]$, which can be estimated by:  
\begin{equation}
r_w[t]=\sum_{k=1}^{M[t]}m_{w,k}[t]\hat{b}_{w,k}[t].
\label{eq:r_w}
\end{equation}
Finally, the CaaS platform\ also keeps track of the total number of active CUs by:
\begin{equation}
N_{\text{tot}}[t]=\sum_{i=1}^{I} p_{i}n_i[t] ,
\label{eq:N_tot}
\end{equation}   
as well as the total compute-unit seconds billed (i.e., already paid to the IaaS provider and available to use) within the   CaaS architecture at any given instant \(t\):
\begin{equation}
c_\text{tot}[t]=\sum_{i=1}^I\sum_{n=1}^{n_i[t]}p_{i}a_{i,n}[t].
\label{eq:c_tot}
\end{equation}

Effectively, $c_\text{tot}[t]$ and $N_\text{tot}[t]$ represent a ``snapshot'' of the compute resources at the $t$th time instant, as they comprise the  available CUSs and CUs which have already been reserved.

The main goals of the proposed CaaS\ platform at every monitoring time instant $t$ are: \textit{(i)} to ensure that each workload \(w\) is executed within its specified TTC, $d_w[t]$, and \textit{(ii)} to match $c_\text{tot}[t]$ to $\sum_{w=1}^{W[t]} r_w[t]$. To this end, the
most critical aspects are: \textit{(i)} defining reliable CUS predictions,  $\hat{b}_{w,k}[t]$, for each data type $k$ within each workload \(w\), \textit{(ii)} confirming the feasibility of each workload's TTC value and selecting the appropriate service rate (i.e., selecting how many CUSs should be allocated to each workload's tasks),   and \textit{(iii)} devising and executing an  algorithm to initialize or terminate CUs according to the demand volume. 

\subsection{Reliable CUS Predictions for Data Types via a Kalman-filter Estimator}
  The platform measures the average  CUSs, \(\tilde{b}_{w,k}\), required for each data type \(k\) of each workload \(w\) running on its instance types, by measuring the time to complete tasks between the previous and the current monitoring instance (\(t-1\) and \(t\)) and refining the measurement. We model this measurement operation mathematically by:
 
\begin{equation}
\forall w,k,t: \tilde{b}_{w,k}[t]=\hat{b}_{w,k} [t]+ v_{w,k}[t],
\label{eq:kalman_measurement}
\end{equation}
where   \(v_{w,k}[t]\) is the measurement noise that deviates  $\tilde{b}_{w,k}[t]$ from the ideal  CUS prediction, $\hat{b}_{w,k} [t]$, at time instant \(t\). We assume that    \(v_{w,k}[t]\)  can be modeled by  independent, identically distributed (i.i.d.), zero-mean Gaussian random variables,  i.e., $\forall w,k:\mathrm{V}_{w,k} \sim \mathcal{N}\left( 0,\sigma_{v}^2 \right)$.

We express the prediction of the required CUSs for each workload and task type at time \(t\) by: 
\begin{equation}
\forall w,k,t: \hat{b}_{w,k}[t]=\hat{b}_{w,k} [t-1]+ z_{w,k}[t],
\label{eq:kalman_estimation}
\end{equation}
with $z_{w,k}[t]$ the process noise \cite{anderson2012optimal}, expressing variability in the execution time of each task type in each workload across time. We assume that   $\forall w,k:z_{w,k}[t]$  can be modelled by  i.i.d., zero-mean Gaussian random variables, i.e., $\forall w,k:\mathrm{Z}_{w,k} \sim \mathcal{N}(0,\sigma_{z}^2)$. Given  \eqref{eq:kalman_measurement} and \eqref{eq:kalman_estimation} and the fact that all noise terms are i.i.d., the noise variances are: $E \{ \mathrm{V}_{w,k}^2 \}  = \sigma_v^2 $, $E\{ \mathrm{Z}_{w,k}^2  \} = \sigma_z^2 $ and the noise covariance is $E\{ \mathrm{V}_{w,k} \mathrm{Z}_{w,k} \}  = 0$. 

For the measurement and prediction model of   \eqref{eq:kalman_measurement} and \eqref{eq:kalman_estimation}, the optimal predictor for  $\hat{b}_{w,k} [t]$   is known to be the Kalman filter \cite{anderson2012optimal}, which provides for the following two time-update equations for our case ($\forall w,k,t$):

\begin{equation}
 \;\pi^-_{w,k}[t]=\pi_{w,k}[t-1]+\sigma_{z}^2, 
\label{eq:kalman_process_update}
\end{equation}

\begin{equation}
\kappa_{w,k}[t]= \frac{\pi^{-}_{w,k}[t]}{\pi^{-}_{w,k}[t]+\sigma_{v}^2}, 
\label{eq:kalman_coeff_update}
\end{equation} 
where $\pi^{-}$ represents the initial update of the process covariance noise $\pi$, and  $\kappa_{w,k}[t]$ is the Kalman gain of the \(k\)th task type of the $w$th workload at time instant $t$. Based on \eqref{eq:kalman_process_update} and \eqref{eq:kalman_coeff_update}, the prediction of  $\hat{b}_{w,k} [t]$   and the noise covariance update can be written as ($\forall w,k,t$):

\begin{equation}
 \hat{b}_{w,k}[t]= \hat{b}_{w,k} [t-1] + \kappa_{w,k}[t] \left( \tilde{b}_{w,k}[t-1] - \hat{b}_{w,k} [t-1] \right), 
\label{eq:kalman_measurement_update}
\end{equation}

\begin{equation}
 \pi_{w,k}[t]=\left ( 1-\kappa_{w,k}[t] \right)\pi^-_{w,k}[t]. 
\label{eq:kalman_process_update2}
\end{equation} 
   
   \textit{Initialization of proposed CUS predictor per workload and task type}:\ For $t=0$ and $\forall w,k$, the platform\ initializes each Kalman-filter estimator with $\tilde{b}_{w,k}[0],$ established via  the initial measurement per workload and input type, and sets: $\hat b_{w,k}[0]=\pi[0]=0$, and $\sigma_z^2=\sigma_v^2=0.5$.

\subsection{TTC Confirmation and Service Rate per Workload} 
Let us assume that a reliable CUS prediction becomes available for workload $w$, $1\leq w\leq W[t_{\text{init}}]$,  at monitoring time instant\footnote{The practical method to determine $t_\text{init}$ is described in Section \ref{Sec5}.} $t_\text{init}$. We can then confirm that $d_w[t_\text{init}]$ (the requested TTC\ for workload $w$ at $t_\text{init}$) is achievable under the appropriate adjustment of the workload \textit{service rate}, $s_w[t]$, for each monitoring time $t$, $t\geq t_\text{init}$. The service rate  $s_w[t]$ corresponds to the number of CUs allocated to workload $w$ for the time interval between monitoring instants $t$ and $t+1$. Fractional values (e.g.,   $s_w[t]=0.7$) indicate that one CU is allocated to workload $w$ for $s_w[t]\times100\%$ of the time between $t$ and $t+1$.  If the combination of  $d_w[t_\text{init}]$  with the workload CUS prediction leads to $s_w[t_\text{init}]>N_{w,\text{max}}$, with $N_{w,\text{max}}$ a predetermined CU upper limit ($\forall w$: $N_{w,\text{max}}=10$ in our experiments),   $d_w[t_\text{init}]$   is extended such that $s_w[t_\text{init}]=N_{w,\text{max}}$. This process confirms $d_w[t_\text{init}]$ (or its extension)\ as the TTC\ for workload $w$. 

The algorithm to determine   $s_w[t]$   for each workload $w$ and each $t\geq t_\text{init}$ is presented in Section \ref{Sec3} and is carried out by the platform based on the predicted CUS per workload.
The CaaS platform selects and executes individual tasks from each workload $w$ according to   $s_w[t]$.

\subsection{Spot Instance Initiation and Termination} 
A direct way to implement the scaling of the required instances is to constantly match the already-billed CUs  [$c_\text{tot}[t]$ of \eqref{eq:c_tot}] to the total CUs required by all workloads ($\sum_{w=1}^{W[t]} r_w[t]$) at each time instant $t$  by initializing or terminating spot instances (a.k.a. ``reactive''\ control \cite{anderson2012optimal}). However, such an approach is not optimal for the following reasons: \textit{(i)}  $\sum_{w=1}^{W[t]} r_w[t]$   depends on the predicted  CUSs to complete the processing of each data type $k$ within each workload $w$; these predictions will not be accurate for all time instants and data types, and this will lead to unnecessary expenditure to initiate and pay for instances that may never be used due to prediction mismatch; \textit{(ii)}  due to the CU billing for large time intervals (e.g., Amazon EC2 spot instances are billed for one hour and GCE instances are billed in 10-minute slots), as well as the associated delay in initialization or termination of instances (in the order of minutes), rapid fluctuations in $\sum_{w=1}^{W[t]} r_w[t]$   (e.g., due to new workloads or workload cancellations by users)\ will cause bursts of initiation or termination requests  and significantly reduce the utilisation rate of the servers, which will result in excessive  bills from the IaaS; \textit{(iii)} without a control mechanism in place to absorb rapid fluctuations in demand, a flurry of spot instance requests may inadvertently cause unwanted spikes in spot instance pricing \cite{song2012optimal}. In the next two sections, we present our\ proposal for best-effort TTC-abiding execution   that ensures proportional fairness amongst all submitted workloads.

\section{Workload Execution with Confirmed TTC}\label{Sec3}

We ensure that each workload is executed within its remaining TTC by an allocation mechanism based on proportional fairness. The proportional fairness goal can then be stated as: at each monitoring instance \(t\) and for each workload $w\:\left( 1 \leq w \leq W[t] \right) $, our platform maximizes an objective function of the service rate, $s_w[t]$, that ensures all workloads are served \textit{proportionally} to their CUS requirement, \(r_{w}[t]\) [given by \eqref{eq:r_w}], and \textit{inversely-proportionally} to their TTC, $d_w[t] $. The latter is defined via an appropriate SLA mechanism once a workload is submitted for execution and an initial workload CUS prediction becomes available. In this work, we adopt the objective function: 

\begin{equation}
f(s_w[t]) = r_{w}[t]\ln(s_w[t]) -d_w[t]s_w[t].
\label{eq:f(p_w)}
\end{equation}

The subtraction in \eqref{eq:f(p_w)} contrasts between the workload's CUS requirement, $r_w[t]$, and the TTC requirement, $d_w[t]$. In addition, following proportional fairness problems in other resource allocation work (notably in cellular network scheduling algorithms \cite{Margolies2014fairness}), we opted for the use of the natural logarithm in the demand side of the objective function and pursue the maximization of  $f(s_w[t])$. Specifically, when the condition  $\sum_{w=1}^{W[t]} r_w[t]    \le c_{\text{tot}}[t]$ is satisfied, it is straightforward to show that the optimal solution to the maximization of \eqref{eq:f(p_w)}  is (\(\forall s_w[t] > 0\))

\begin{equation}
s_w^*[t] = \text{arg} \; \text{max} \left\{ f \left( s_w[t] \right) \right\}=\frac{r_{w}[t]}{d_w[t]}.   
\label{eq:arg_max}
\end{equation}
This corresponds to the case where enough CUs are available to accommodate the demand and, therefore, allocation of service rates is carried out according to the required CUSs and TTC per workload at each monitoring time instant \(t\). We can then calculate the total required CUs for optimal operation as: 

\begin{equation}
N_\text{tot}^{*}[t]=\sum _{w=1}^{W[t]}s_{w}^*[t]=\sum _{w=1}^{W[t]}\frac{r_{w}[t]}{d_w[t]}.   
\label{eq:N_tot_optimal}
\end{equation}

However, due to volatility in both workload submission and CU availability in the CaaS platform, it is likely that, for most monitoring instances $t$,   $N_\text{tot}^{*}[t]$ differs from $N_\text{tot}[t]$ [the actual number of CUs, calculated by \eqref{eq:N_tot}]. In such cases, we can  adjust the optimal service rates of   \eqref{eq:arg_max} proportionally to the relative distance between   $N_\text{tot}^{*}[t]$ and $N_\text{tot}[t]$. Specifically, if $N_\text{tot}^{*}[t]>N_\text{tot}[t]+\alpha,$ with $\alpha$ the AIMD additive constant defined in the next section ($\alpha > 0$), we downscale the optimal service rate of each workload to:\

\begin{eqnarray}
\forall w: s^{-}_w[t] & = &\frac{r_{w}[t]}{d_w[t]} \left( 1-\frac{N_\text{tot}^{*}[t]-N_\text{tot}[t]-\alpha}{N_\text{tot}^{*}[t]}  \right) \nonumber \\ & = &\frac{ N_\text{tot}[t] + \alpha }{N_\text{tot}^{*}[t]} s_w^*[t] .   
\label{eq:s_w_scaledown}
\end{eqnarray}
If  $N_\text{tot}^{*}[t]<\beta N_\text{tot}[t]$, with $\beta$ the AIMD\ scaling constant defined in the next section ($0<\beta<1$), we upscale the optimal service rate of each workload  to:

\begin{eqnarray}
\forall w: s^{+}_w[t] & = & \frac{r_{w}[t]}{d_w[t]} \left( 1+\frac{\beta N_\text{tot}[t]-N_\text{tot}^{*}[t]}{N_\text{tot}^{*}[t]}  \right) \nonumber \\ & = & \frac{\beta  N_\text{tot}[t]}{N_\text{tot}^{*}[t]}s_w^*[t] .   
\label{eq:s_w_scaleup}
\end{eqnarray}
Finally, if $\beta N_\text{tot}[t] \leq N_\text{tot}^{*}[t] \leq N_\text{tot}[t]+\alpha$,  the service rates of    \eqref{eq:arg_max} are used. The use of $\alpha$ and  $\beta$ in \eqref{eq:s_w_scaledown} and \eqref{eq:s_w_scaleup} ensures the service rate adjustment is considering the possible additive increase or multiplicative decrease that may occur via the AIMD\ algorithm after the service rate allocation is established for the interval between $t$ and $t+1$.\ 

%
%

  \section{Scaling with Additive Increase Multiplicative Decrease}\label{Sec4}

For any CaaS system, $N_\text{tot}^{*}[t]$ of \eqref{eq:N_tot_optimal} and $N_\text{tot}[t]$ of \eqref{eq:N_tot}   must be tightly coupled in order to ensure that the available compute-unit time can meet the service demand and TTC requirements at any instant. This is because, if $N_\text{tot}^{*}[t]$ is substantially higher than  $N_\text{tot}[t]$, the delay to complete pending workloads can increase substantially and workload TTCs may be violated. Conversely, when $N_\text{tot}^{*}[t]$ is  significantly smaller than  $N_\text{tot}[t]$,   unnecessary resource costs may occur. Therefore, and in conjunction with the fact that  billing comes in hourly increments in Amazon EC2 spot-instances, sudden surges or dips in demand will have a detrimental effect in the delay or cost of the deployment of platform. Hence, the goal of the platform is to maintain the resource reservation and workload service rates at the correct level. To this end, we propose the AIMD algorithm of  Fig. \ref{fig:FigAIMD}. By controlling the additive and scaling constants, $\alpha$ and $\beta$ respectively, we can examine the behavior of the platform under a wide variety of workload submissions. It should be noted that the corresponding problem of selecting which spot instances to terminate  in the event that $N_\text{tot}[t] > N_\text{tot}^*[t]$ is trivial: per instance type, the prudent action is always to terminate spot instances with the smallest remaining time before renewal.

\begin{figure}[t]
1 \ \textcolor[rgb]{0,0.501961,0.501961}{\% algorithm iterations for any monitoring time instant $t$} \

2 \ \textbf{if } $N_\text{tot}[t] \leq N_\text{tot}^{*}[t]$    \\
3 \ \ \ \ \texttt{incr} $ = $ \texttt{TRUE}\\ 
4  \ \textbf{else }    \\
5 \ \ \ \ \texttt{incr} $ = $ \texttt{FALSE}\\ 
6 \ \textcolor[rgb]{0,0.501961,0.501961}{\% application of AIMD to tune $N_\text{tot}$ for the next instant} \\
7 \ \textbf{if } \texttt{incr == TRUE} \\
8 \ \ \ \ $N_\text{tot}[t+1]= \min\{N_\text{tot}[t] + \alpha, N_\text{max}\}$ \textcolor[rgb]{0,0.501961,0.501961}{\% add more CUs}  \\
9 \ \textbf{else} \\
10 \ \ $N_\text{tot \\}[t+1] = \max\{\beta N_\text{tot}[t], N_\text{min}$\} \textcolor[rgb]{0,0.501961,0.501961}{\% remove CUs} \\
 
\caption{Proposed AIMD algorithm; $\alpha$ is a positive constant, $\beta$ is a constant such that $0 < \beta \le 1$,  and $N_\text{max}$ and $N_\text{min}$  are the upper and lower bounds for $N_\text{tot}[t]$.}
\label{fig:FigAIMD}
\end{figure}

We refer to the  work of Shorten \textit{et. al.} \cite{shorten2006positive} for details on the the stability and convergence properties of AIMD algorithms. A key aspect from their analysis is that fast convergence to an equilibrium state is achieved if $\beta$ is small and smoother transitions are expected if $\beta$ is close to unity  \cite{shorten2006positive}. After extensive experimentation, we opted for the values of   $\beta=0.9$ and $\alpha=5$,  which exhibit sufficiently-fast convergence while at the same time ensuring that CUs are not released prematurely. 

While the AIMD\ algorithm tunes the total CU value, $N_\text{tot}[t]$, it does not select which instance types to deploy out of the $I$ possible. As detailed in Appendix \ref{sec:Appendix}, the recent status of Amazon spot instance pricing provides for proportional increase of pricing according to the number of compute units per instance. Moreover, the single-CU instance type exhibits the minimum price volatility, thereby making it the safest instance type to use. Therefore, we opt to use only single-CU instances in our experiments, i.e., $I=1$ and $p_1=1$, which alleviates the problem of selecting amongst a variety of instance types. However, depending on the evolution of pricing data from the IaaS provider, future work will expand our results into a variety of instance types.

\section{Experiments}\label{Sec5}

In order to examine our proposals, we have deployed our CaaS platform using single-CU \texttt{m3.medium} spot instances of Amazon EC2 (see Appendix  \ref{sec:Appendix} for more details). Each instance has a corresponding component that requests   new tasks to process once it 
detects that pending workloads are available with non-zero service rates. In addition, one reserved EC2 instance, serving as the ``estimator'' component of the platform, calculates the Kalman filter prediction estimates based on the CUS measurements per task.    Under predetermined TTC per workload (which is confirmed by platform after an initial CUS prediction becomes available for the workload), it then derives the service rate per workload  in fixed time periods (i.e., within 1--5 minute intervals), as described in Section \ref{Sec3}. This is communicated to the other instances. The estimator component also carries out the AIMD algorithm of Section \ref{Sec4} in order to control the increase or decrease of spot instances according to the demand. The utilized AIMD parameters for all experiments were set to: $\alpha=5$,  $\beta=0.9$, $N_\text{min}=10$, $N_\text{max}=100$ and $\forall w$: $N_{w,\text{max}}=10$ (maximum service rate per workload). 
\subsection{Utilized Workloads}

\begin{figure}[t]
\begin{center}
\includegraphics[scale=0.30]{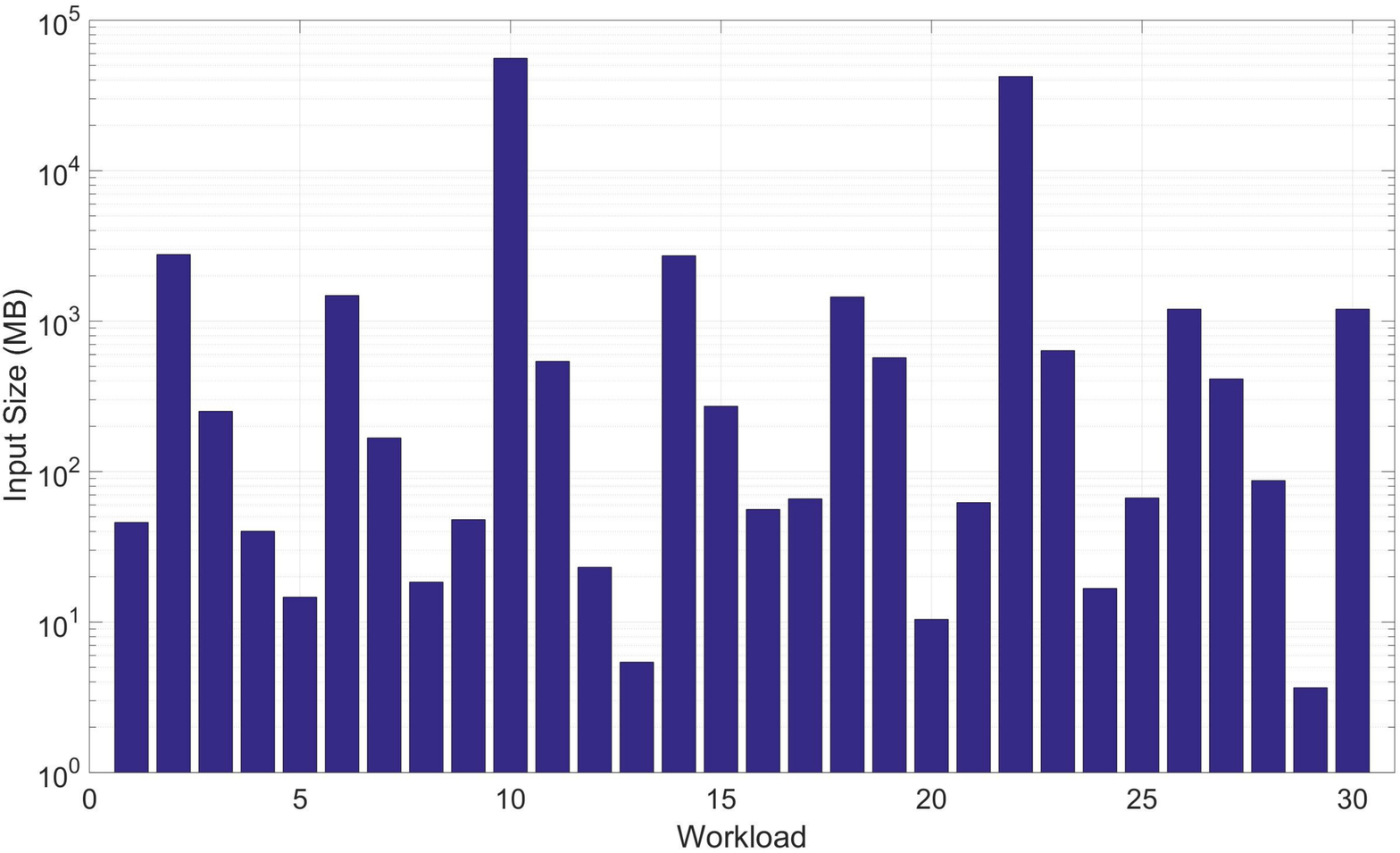}
\end{center}
\caption{Size of inputs for each of the thirty workloads used in our experiments.}
\label{fig:FigWorkloads}
\end{figure}

 Thirty different workloads, each with a  random number of tasks were used in our experiments. Eight of the workloads were scripts running the Viola-Jones  classifier \cite{viola2004robust} for face detection in  images. The range of possible values for the number of inputs (i.e., images or videos) for these workloads
 was between 1 and 1000. Eight of the workloads were scripts using FFMPEG to transcode videos to different bitrates via a variety of codecs \cite{kontorinis2009statistical,andreopoulos2007adaptive,garcia2010study}. Each workload had between 1 and 20 videos to transcode, and we also added two large transcoding workloads with 200 and 300 videos. These were used to examine
 the responsiveness of the platform under sudden spikes of demand. Seven of the workloads were using the OpenCV BRISK keypoint detector and descriptor extractor \cite{ leutenegger2011brisk}. Finally, seven workloads used the Scale Invariant Feature Transform (SIFT) salient point descriptor \cite{lowe2004sift}, which was deployed as compiled Matlab code with the Mathworks \texttt{deploytool}. These were selected as representative examples for image processing operations, amongst a larger pool of image and video processing experiments using wavelet transforms \cite{andreopoulos2002new,andreopoulos2001analysis,andreopoulos2001local} and other codecs \cite{andreopoulos2000hybrid,kontorinis2009statistical,andreopoulos2007adaptive}. The total size of the inputs per workload is given in Fig. \ref{fig:FigWorkloads}.
Workloads were introduced once every five
 minutes in the order depicted in Figure \ref{fig:FigWorkloads}.

\subsection{Performance of Kalman-based CUS Prediction }

The proposed Kalman-based CUS\ prediction process of Section \ref{Sec3} is compared against the  ``ad-hoc'' estimator that carries out the CUS\ prediction of \eqref{eq:kalman_measurement_update}, albeit with the scaling coefficient being set to the fixed value: $\kappa_{w,k}[t]=0.1$, which was shown to perform best amongst other settings. Moreover, as an external comparison, we also utilize the well-known second-order autoregressive moving average (ARMA) estimator of Roy \textit{et. al.}  \cite{roy2011efficient}  that has been shown to perform well for workload  forecasting. ARMA predicts the CUS\ required to complete a workload at time $t+1$ via  
\begin{eqnarray}
\hat{b}_{w,k} [t+1] & = & \delta \times b_{\text{norm},w,k} [t]+ \gamma \times  b_{\text{norm},w,k} [t-1] \nonumber\\
& + & (1-\delta-\gamma)  \times  b_{\text{norm},w,k} [t-2],   
\label{eq:ARMA_CUS_pred}
\end{eqnarray}
where: $b_{\text{norm},w,k} [t,t-1,t-2]$ are calculated by summing the total execution time of data type $k$ of workload $w$ at times $t,t-1,t-2$ and dividing it by the percentage of the workload that has been completed until then; and $\delta$ and $\gamma$ are scalars having the values recommended by  Roy \textit{et. al.}  \cite{roy2011efficient}. ARMA was chosen as the most suitable benchmark as other workload forecasting methods (like  the ARIMA model   \cite{debusschere2012hourly,Calheiros2014arima}) require extensive past measurements from previous executions of other workloads, as well as a long sequence of measurements in order to produce reliable prediction estimates, thereby making them unsuitable in our case.  

\begin{figure}[t]
\begin{center}
\includegraphics[scale=0.31]{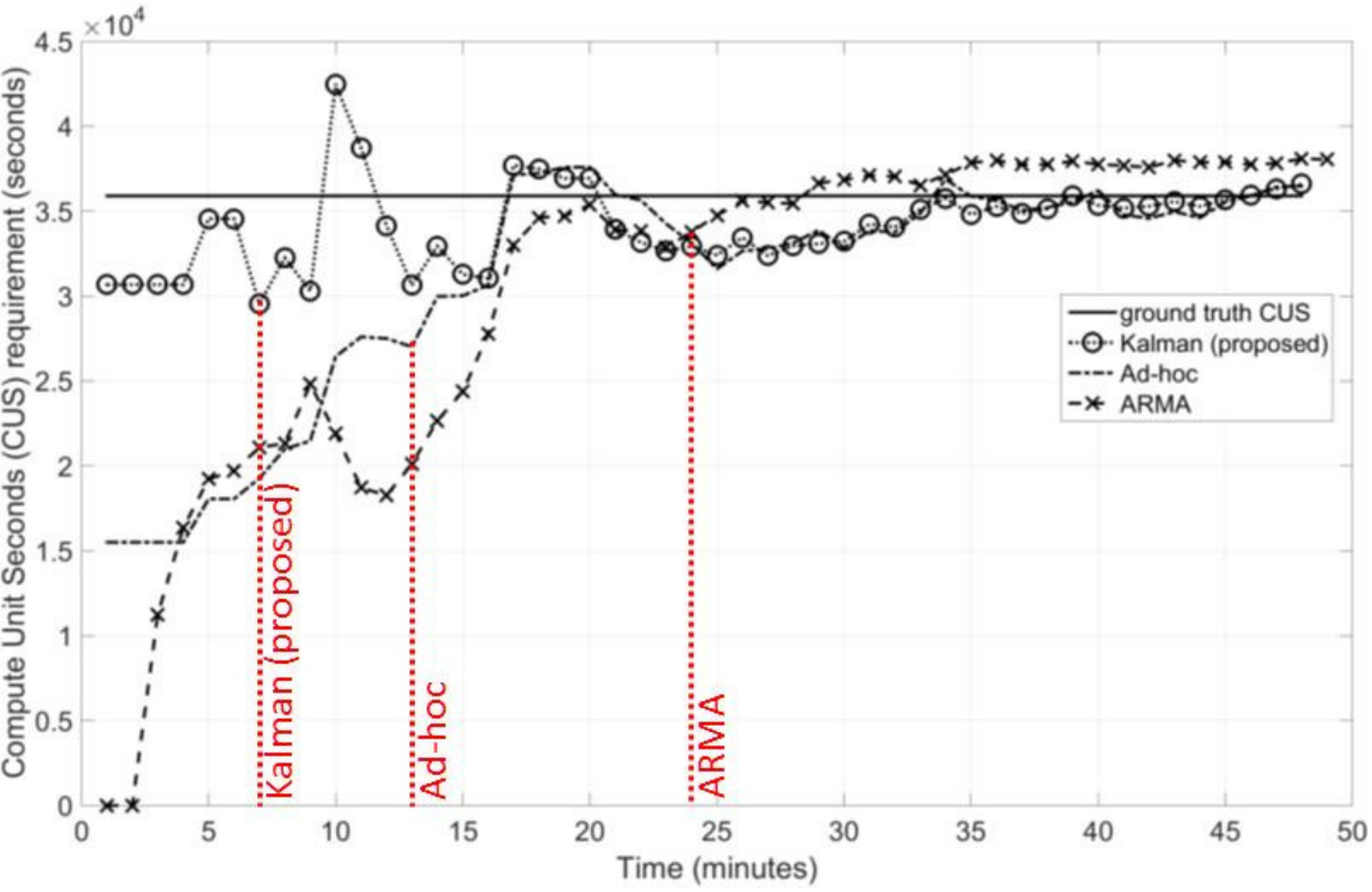}
\end{center}
\caption{Example of the convergence of various CUS prediction methods for the case of an FFMPEG\ workload under 1-min monitoring interval.   }
\label{fig:EstimatorsSlope}
\end{figure}

A representative example of the convergence behaviors of all methods under comparison is given in Fig. \ref{fig:EstimatorsSlope}.  As illustrated in the figure, the Kalman and ad-hoc estimator exhibit an underdamped behavior until convergence.  We can therefore use the slope of the CUS\ prediction across time to determine the monitoring time instant $t_\text{init}$ when the proposed  Kalman and the ad-hoc estimator can provide a reliable CUS\ prediction per workload and task type. Specifically, when the slope of the CUS\ prediction becomes negative for the first time, each estimator establishes a CUS\ prediction for each workload with acceptable accuracy.  However, ARMA\ does not exhibit such underdamped behavior, since it is a moving-average based estimator. Therefore, we relied on a conventional convergence detection criterion for ARMA: when the ARMA\ prediction value deviation within the window of the last three measurements is found not to exceed 20\% from the mean value derived from the values of the window (ten measurements are used for the case of 1-min monitoring), we determine that the prediction is reliable enough to be used. The setup for the window size and variability threshold was selected after testing with a variety of possible values. In the example of  Fig. \ref{fig:EstimatorsSlope}, the time instant when each method reaches its reliable prediction under the described setup is marked with the red dotted vertical line.   

Table \ref{tab:TabTTCest} presents the average time each estimator took to reach its CUS prediction for each workload type, as well as the CUS\ percentile mean absolute error (MAE). The summary over all workloads (per monitoring interval) is given at the bottom of the table. Evidently, the proposed Kalman-based approach reduces the average time to reach a reliable prediction by more than 20\% in comparison to the other estimators and is found to be the quickest estimator in all but one case. At the same time, the proposed estimator attains comparable  accuracy to the ad-hoc estimator and is found to be significantly superior to ARMA. This is especially pronounced in the case of the 1-minute monitoring, where the use of the proposed Kalman-based approach instead of an ARMA approach provides for 38\% reduction in prediction time and decreases the average prediction error from 16.4\% to 4.5\%. This indicates that, under the usage of the proposed CUS\ estimator and 1-minute monitoring, the GCI\ is expected to have reliable predictions per workload (and thereby confirm that its requested TTC is achievable) within 6--11 minutes from its launch. 
Finally, when we compare the performance of one-minute monitoring to five-minute monitoring, Table \ref{tab:TabTTCest} shows that increase in the measurement granularity results in significant improvement in both the accuracy and time required to reach a reliable prediction. Specifically, for the proposed Kalman estimator, increased monitoring frequency reduces the the average prediction time by 44\% and reduces the overall MAE from 13.1\% to 4.5\%.

\ 


\begin{table*}
\caption{Average time to reach CUS  prediction per type of workload and percentile Mean Absolute Error (MAE)\ of the derived prediction. The last column presents the percentile time reduction when switching from 5-min monitoring to 1-min monitoring intervals. The best result per category is indicated in boldface font.}

\begin{center}
\begin{tabular}{|c|c|c|c|c|c|}
\hline 
Control interval & \multicolumn{2}{c|}{5-min monitoring } & \multicolumn{2}{c|}{1-min monitoring }& Time 

Reduction (\%) by going \tabularnewline
 
\textbf{ Face Detection} &  Time & MAE (\%)\ & Time & MAE (\%) &  from 5-min to 1-min monitoring\tabularnewline
\hline 
\hline 
Kalman-based & \textbf{13m 45s} & 5.6 & \textbf{10m 38s} & \textbf{4.6} & 22.7\textbf{}   \tabularnewline
\hline 
Ad-hoc  & 28m 08s & \textbf{4.5} & 17m 53s  & 5.3 & 36.4  \tabularnewline
\hline
 ARMA & 23m 08s & 22.1 & 12m 08s & 27.8 & 47.6\tabularnewline
\hline
\hline 
\textbf{Transcoding} & Time & MAE (\%) &Time & MAE (\%) & \tabularnewline
\hline 
\hline 
Kalman-based & \textbf{16m 53s} & 14 & \textbf{07m 54s}  & 7.8 & 53.2    \tabularnewline
\hline 
Ad-hoc  & 26m 53s & \textbf{8.9} &  10m 36s  & \textbf{1.5} & 60.6   \tabularnewline
\hline 
ARMA & 28m 08s & 13.9 & 18m 45s & 18.1 & 33.4  \tabularnewline
\hline
\hline 
\textbf{Feat. Extraction} & Time & MAE (\%) &Time & MAE (\%) &  \tabularnewline
\hline 
\hline 
Kalman-based & \textbf{13m 34s} & 12.1 & 11m  54s&  \textbf{1.4} & 12.3 \ \tabularnewline
\hline 
Ad-hoc  & 18m 34s & 6.4 & 20m 24s  &1.9 &-9.9  \tabularnewline
\hline
ARMA & 20m 43s & \textbf{5.7} & \textbf{11m 09s} & 12.1 & 46.2 \tabularnewline
\hline
\hline 
\textbf{SIFT} & Time & MAE (\%) & Time &MAE (\%) & \tabularnewline
\hline 
\hline 
Kalman-based & \textbf{21m 26s} & 20.6  & \textbf{06m 18s} & 4.1 & 70.6  \tabularnewline
\hline 
Ad-hoc  & 23m 54s & 18.9 & 08m 06s & \textbf{0.1} & 66.1 \tabularnewline
\hline
ARMA & 20m 00s & \textbf{20.1}  & 15m 00s & 7.6 & 25.0\tabularnewline
\hline
\hline 
\textbf{Overall Average} & Time & MAE (\%) & Time & MAE (\%) & \tabularnewline
\hline 
\hline 
Kalman-based & \textbf{16m 25s} & 13.1  & \textbf{09m 11s} & 4.5 & 44.1  \tabularnewline
\hline 
Ad-hoc  & 24m  22s & \textbf{9.7} & 14m 15s & \textbf{2.2} & 34.6 \tabularnewline
\hline 
ARMA & 23m 00s & 15.5 & 14m 15s & 16.4 & 38.0\tabularnewline
\hline

\end{tabular}
\end{center}
\label{tab:TabTTCest}
\end{table*}

\subsection{Results for Cumulative Cost of Workload Execution }

We now investigate the management of spot instances so that each workload is completed under a fixed TTC that is sufficiently large to allow for fluctuation in the number of utilized instances.  

As external comparisons, our first choice is Amazon's Autoscale service (termed as ``Amazon AS''), which is widely deployed in practice  \cite{tighe2014autoscale}. Amazon AS does not carry out CUS prediction or TTC-abiding execution, and one can  only control the number of instances based on CPU utilization and bandwidth constraints. Therefore, under these conditions, we configured all workloads to execute within an Amazon AS group that examines the average CPU usage at all utilized CUs in five-minute intervals. If the group detected that the average CPU utilization was more than 20\%, new instances were started\footnote{After extensive experimentation, the value of 20\% was found to provide for the best results with Amazon AS. This is because average\ utilization values between 18\%\ and 22\% represent the average CPU\ usage observed within active time intervals when an instance alternates between downloading files (2\%--10\%\ CPU\ utilization) and actually  executing a compute-intensive task (close to 100\% CPU\ utilization).}. Otherwise, Amazon AS terminated some of the active instances. We then executed all workloads in Amazon AS and measured the longest time to complete a workload under two scaling policies. The first represented a conservative approach where reducing the execution time is not of critical importance. In this case, a single instance is added or removed when a monitoring interval occurs. The longest completion time was found to be 2 hr 7min. The second scaling policy started and stopped ten instances instead of one, to represent a scenario where reduced execution time is of importance. In this case, the longest time to complete a workload was found to be 1 hr and 37 min. Both of these times were then used as the two fixed TTC settings for all workloads in our platform. 

Beyond Amazon AS, in order to benchmark our AIMD-based scaling of Fig. \ref{fig:FigAIMD} against other alternatives for CU adjustment, we utilized  the mean-weighted-average and linear-regression methods   of Gandhi, Krioukov \textit{et. al.}  \cite{Gandhi2012autoscale,krioukov2011napsac} (termed ``MWA'' and ``LR'', respectively) to set the number of CUs for the next monitoring interval, $N_\text{tot}[t+1]$. We selected MWA and LR for our comparisons because previous work  \cite{Gandhi2012autoscale} has shown them to be amongst the most accurate predictive resource controllers. Both MWA and LR utilized the proposed Kalman-based CUS prediction process and the service rate allocation of  \eqref{eq:N_tot_optimal} to determine when to increase or decrease CUs. Specifically: \textit{(i)} MWA\ sets the number of CUs via

\begin{equation}
N_\text{tot}[t+1]=\frac{1}{6} \sum_{i=t-5}^{t}N_\text{tot}^{*}[i],   
\label{eq:MWA_CUs}
\end{equation}
where $N^*_\text{tot}$ is the optimal number of CUs derived via \eqref{eq:N_tot_optimal} for each monitoring time instant; \textit{(ii)} LR\ sets $N_\text{tot}[t+1]$ to be the result of extrapolating the line derived via linear regression from $\{N_\text{tot}^*[t], \dots, N_\text{tot}^*[t-5]  \}$ (current plus five previous CU settings). Finally, in order to see the performance of the direct-compensation approach, we also utilized the case where no filtering or other adjustment is being used and we simply set  $N_\text{tot}[t+1]=N_{\text{tot}}^{*}$   (termed as ``Reactive'').


\begin{figure}[t]
\begin{center}
\includegraphics[scale=0.31]{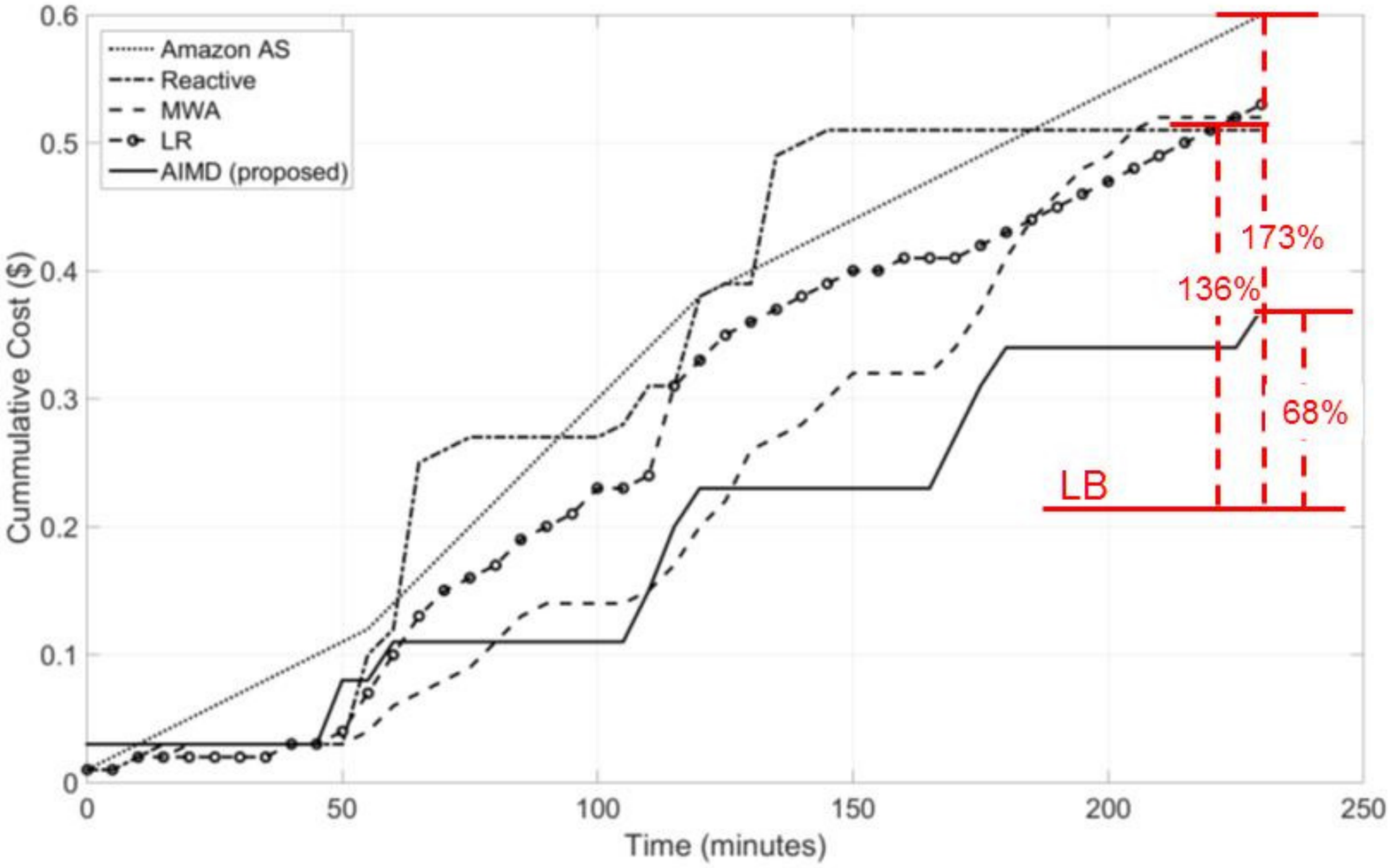}
\end{center}
\caption{Cumulative cost of processing all workloads of Fig. \ref{fig:FigWorkloads} under fixed TTC of  2 hr 7min per workload.}
\label{fig:Cost}
\end{figure}

\begin{table*}[t]
\caption{Summary of overall cost of different methods as well as comparisons with the proposed method and the lower bound }

\begin{center}
\begin{tabular}{|c|c|c|c|c|c|c|}
\hline
System & AIMD (proposed)  & Reactive & MWA & LR & AS & LB \tabularnewline  
\hline
Overall cost (\$) & 0.41 & 0.51 & 0.52 & 0.53 & 1.02 & 0.22 \tabularnewline \hline
Average cost reduction of proposed vs. other methods (\%) & -- & 20 & 21 & 23 & 60 & -- \tabularnewline
\hline
Average cost increase above LB (\%)\ & 86 & 132 & 136 & 141 & 364 & -- \tabularnewline
\hline 
Max. \#\ of instances at any time by each method & 13 & 28 & 21 & 24 & 91 &-- \tabularnewline \hline
\end{tabular}
\end{center}
\label{tab:TabCostSummary}
\end{table*}

\begin{figure}[t]
\begin{center}
\includegraphics[scale=0.31]{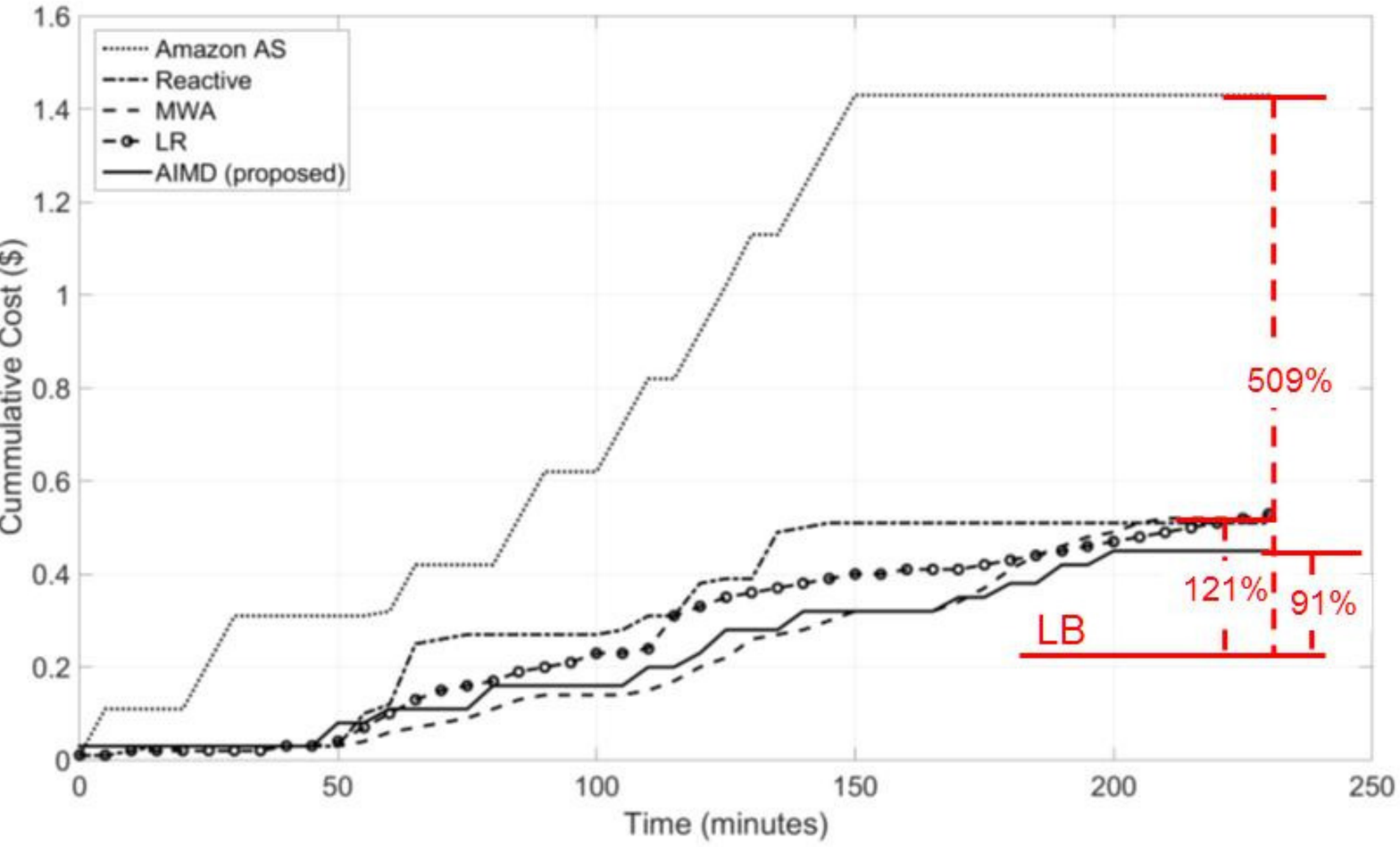}
\end{center}
\caption{Cumulative cost of processing all workloads of Fig. \ref{fig:FigWorkloads} under fixed TTC of  1 hr 37min per workload.}
\label{fig:CostLowDelay}
\end{figure}

Figure \ref{fig:Cost} and Figure \ref{fig:CostLowDelay} show the cumulative cost of each approach during the course of both experiments with the two TTC values. Evidently, the cost of Amazon AS is  significantly higher than that of all other approaches. This is primarily because the Amazon AS is the only approach that does not use CUS predictions and instead bases its decisions solely on CPU\ utilization. Therefore, it continues to scale up the number of instances even when it is nearing completion of the workloads' processing and only scales down after workloads have been completed and CPU utilization decreases due to inactivity.   

Amongst MWA, LR and Reactive, MWA is superior as it incurs less cost for the majority of the experiment (and, as expected, Reactive is the worst). However, all three methods end up incurring very comparable cost for the completion of all workloads. Interestingly, Reactive turns out to be (marginally)\ the cheapest of the three for this experiment even though it uses the largest number of instances of the three methods at one point. The reason for this is that, while Reactive scales up very quickly it also scales down rapidly and, for this particular experiment, this behaviour worked in its favour. However, this is not expected to be always the case, as Reactive does leave many instances  idle for a large portion of their billed time.

 The proposed AIMD-based scaling initially scales up when it detects the large workloads, then maintains this level, and then begins to scale down as it nears the experiment completion. For the experiments of Figure \ref{fig:Cost}, this leads to overall savings of 30\% against MWA, 29\% against LR, 27\% against\ Reactive and 38\%\ against Amazon AS. For the experiments of Figure \ref{fig:CostLowDelay}, the equivalent savings were:\  14\%, 15\%, 12\%\ and 69\%. Overall, beyond the advantage of providing for scaled-up execution under TTC constraints, the 38\%--69\% savings demonstrated in Figure \ref{fig:Cost} and Figure \ref{fig:CostLowDelay}  allow for significant profit margin for cloud\ service providers that would deploy large-scale multimedia applications via the techniques used in our platform, versus utilizing Amazon AS\ directly. 

The overall savings for both experiments, as well as the maximum number of instances used by the proposed algorithm against all other benchmarks are summarized in  Table \ref{tab:TabCostSummary}. It should be emphasized that, beyond the cost savings, all the workloads in the proposed AIMD approach finished before their execution time exceeded the predetermined TTC of each experiment. Such TTC-abiding execution is a significant feature that Amazon AS cannot provide.

Finally, the bottom right of Figure \ref{fig:Cost} and Figure \ref{fig:CostLowDelay} includes a red horizontal line indicating the estimated billing if all workloads would be processed such that all billed instances would be occupied 100\%\ of the time. This constitutes the lower bound for the billing cost (termed ``LB'') as no operational approach can achieve lower cost. Evidently, the proposed approach incurs 68\%--91\%\ higher cost than LB, but all other approaches incur 135\%--510\%\ higher cost than LB. This demonstrates that the proposed AIMD-based scaling of CUs is a simple and effective method towards approaching the lowest possible cost in the cloud computing infrastructure while at the same time satisfying the TTC\ constraint of each workload. \ \

\subsection{Comparison Against Amazon Lambda}

Recently, Amazon begun offering its own CaaS service for the execution of Javascript code via its Lambda service. Despite this being more limiting due to the inefficiency of Javascript code, we compared the cost of running three large Javascript-based workloads on our platform and Lambda. In this experiment we ran ``blur'', ``rotate'' and ``convolve'' operations from the Javascript version of the widely-used  ImageMagick image manipulation program \cite{imagemagick}. We chose these functions as they represent a cross section of computational requirements of the various ImageMagick functions. Each function was executed on 25,000 images encompassing a wide variety of sizes and pixel counts. We also opted for the 1024MB-memory configuration for all Lambda functions to avoid any memory bottlenecks during execution. Again, our platform was tuned to match the execution time of each workload in Lambda. This was done because the latter is dependent on how quickly requests can be sent to call the functions through the Amazon Web Service Command Line interface (or any other such API), while the execution time for workloads in our platform is completely tunable based on their specified TTC. This flexibility of TTC-abiding execution\ per workload is an advantage of our proposal against Lambda.

A comparison of the cost of executing the workloads is given in Table \ref{tab:LambdaComparison}. It is interesting to notice that, as the run time of the function decreases, Lambda becomes a more viable option. For example,  the average cost of running the most compute-intensive function (Blur function in Table \ref{tab:LambdaComparison})  was 3.34 times higher on Lambda than it was on our platform. In contrast the average cost of running the fastest and least compute-intensive function (the rotate function) was found to be slightly less on Lambda than on our platform. Overall, we were able to run the workloads on our platform at more than 2.5 times lower cost (60\% reduction)\ in comparison to Amazon Lambda. This provides for substantial profit margin for a cloud service provider to deploy a large-scale multimedia application via the proposed approach instead of Lambda.

\begin{table}
\caption{Average cost of ImageMagick functions per image of the  25,000 dataset for Proposed platform and Amazon's lambda. }

\begin{center}
\begin{tabular}{|c|c|p{2.5cm}|c|}
\hline
Function & Lambda Cost (\$) & Proposed Platform Cost (\$) & Ratio \tabularnewline
\hline
Blur & $4.74 \times 10^{-5}$ & $1.42 \times 10^{-5}$ & 3.34\ \tabularnewline
\hline
Convolve & $1.68 \times 10^{-5}$ & $6.05 \times 10^{-6}$ & 2.78 \tabularnewline
\hline
Rotate & $5.5 \times 10^{-6}$ & $6.8 \times 10^{-6}$ & 0.81 \tabularnewline
\hline
Overall Average & $2.32 \times 10^{-5}$ & $9.20  \times 10^{-6}$ & 2.52 \tabularnewline
\hline

 \end{tabular}
\end{center}
\label{tab:LambdaComparison}
\end{table}

\section{Conclusions}\label{Sec6}

We propose and explore three novel aspects for resource management and prediction in Computation-as-a-Service (CaaS)\ frameworks: \textit{(i)} a resource allocation algorithm based on proportional fairness; \textit{(ii)}  the allocation or termination of compute units based on the Additive Increase Multiplicative Decrease (AIMD)\ algorithm; \textit{(iii)}\ the prediction of compute-unit seconds for each type of task with each executed workload via a Kalman-based estimator. Experiments based on Amazon EC2 spot instances demonstrate that, unlike all existing Platform-as-a-Service and Software-as-a-Service frameworks, our platform provides for extreme scaling of  computing tasks (like large-scale transcoding, face detection and feature extraction workloads), without requiring any modification in the users' code base, and at substantially-reduced cost against all other alternatives. In addition, our platform allows for execution under time-to-completion constraints, unlike other platforms. The baseline form of the proposed CaaS framework is available{} at http://www.dithen.com.\

\appendices 

\section{\label{sec:Appendix}}

We briefly analyze the computation costs of Linux instances on AWS EC2, as EC2 is considered to be the largest public cloud service provider today  \cite{Choy2014AWSLargest}  and our system is tested and deployed on the EC2 infrastructure. A comparison of the cost and EC2 compute units (ECUs) of various instance types is given in\footnote{Table \ref{tab:CostComparison} does not include all instance types available on Amazon's EC2. However, all non-included instances are memory, computation or storage variants of the instances depicted in Table \ref{tab:CostComparison}.} Table \ref{tab:CostComparison}. An ECU is defined as ``equivalent CPU capacity of a 1.0-1.2 GHz 2007 Opteron or 2007 Xeon processor''. The  \texttt{m3.medium}  instance (utilized in this paper)\ is a single CU instance with clock speed of 3.0--3.6GHz. From the table we can also see that the larger instances consist of increasing numbers of CUs (i.e., virtual cores available for computations) with similar clock speeds. We can also see that the ``On Demand'' cost and spot prices are both linearly-dependent on  the number of CUs. Thus, we can conclude that it is more efficient to use a large number of cheaper instances than small number of more expensive instances, as it allows for greater granularity when controlling the number of active instances without any corresponding increase in cost.

\begin{table*}
\caption{Cost of Various Linux Instances on the Amazon EC2 Platform in the North Virginia Region }

\begin{center}
\begin{tabular}{|p{4.5cm}|c|c|c|c|c|c|}
\hline
\textbf{Instance Type} & m3.medium & m3.large & m3.xlarge & m3.2xlarge & m4.4xlarge & m4.10xlarge\tabularnewline
\hline
\hline
EC2 compute units (ECUs) & 3  & 6.5 & 13 & 26 & 53.5 & 124.5\tabularnewline
\hline
CUs  & 1 & 2 & 4 & 8 & 16 & 40\tabularnewline
\hline
On-demand cost (\$) & 0.067  & 0.133 & 0.266 & 0.532 & 1.008 & 2.52\tabularnewline
\hline
Spot  price (\$)\ & 0.0081 & 0.0173 & 0.0333 & 0.066 & 0.1097 & 0.5655 \tabularnewline
\hline
Cost reduction when using spot  (\%)& 88 & 87 & 87 & 88 & 89 & 78  \tabularnewline
\hline

 \end{tabular}
\end{center}
\label{tab:CostComparison}
\end{table*}
\begin{figure}[t]

\includegraphics[scale=0.7]{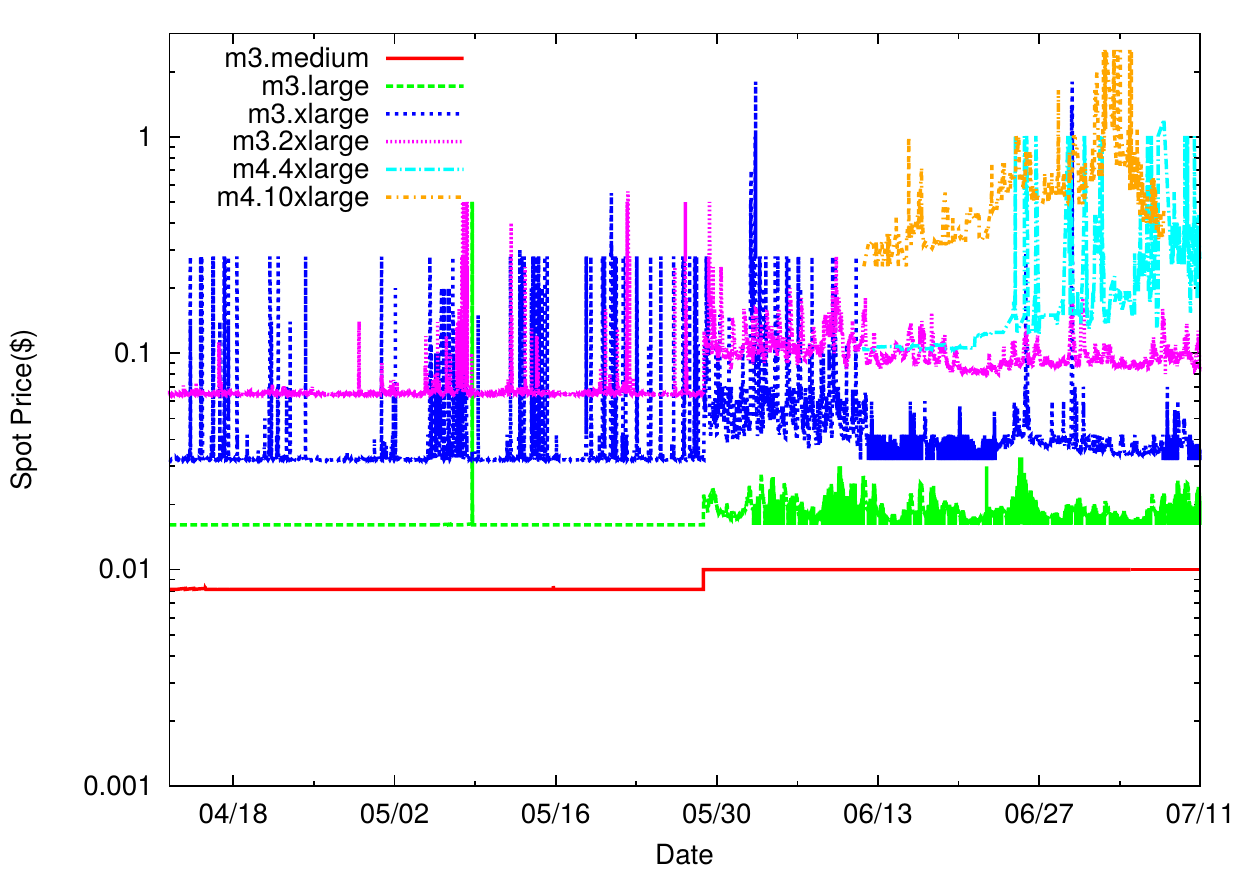}
\centering

\caption{Spot Price for various instance types from 11\textsuperscript{th} of April to the 11\textsuperscript{th} July 2015 }
\label{fig:FigSpot}
\end{figure}

From Table \ref{tab:CostComparison} we can also see the difference between the ``On Demand'' cost and the Spot price\footnote{The Spot prices depicted in Table \ref{tab:CostComparison} were taken on the 10\textsuperscript{th} July 2015.}. Spot instances are instances that will only function when a user's bid is greater than the current spot price. Essentially, the user gives up certainty of having computational resources available, in exchange for a significant reduction in the cost. We can see from Table \ref{tab:CostComparison} that this reduction ranges from 78\% to 89\%. However, it is difficult to run a CaaS service without guarantees of the availability of computational resources, so an analysis of the fluctuation of the spot price is necessary to determine if spot instances should be utilized.

The spot instance price for various instance types in the three-month period from the 11\textsuperscript{th} of April to the 11\textsuperscript{th} July 2015 is shown in\footnote{In the case of \texttt{m4.4xlarge} and \texttt{m4.10xlarge} data was only available from the 11\textsuperscript{th} June.} Figure \ref{fig:FigSpot}. Evidently, the volatility of the spot price is proportional to number of CUs that an instance possesses. Therefore, while it would be difficult to rely on a \texttt{m4.10xlarge} spot instance, the spot price of the \texttt{m3.medium} spot instance is remarkably stable. Specifically, at no point in the three month period does the \texttt{m3.medium} spot price exceed \$$0.01$. Therefore, we can conclude that a significant reduction in cost can be achieved by using \texttt{m3.medium} spot instance with little effect on the reliability of the service, and with more flexibility than when using larger spot instances with more CUs.

\bibliographystyle{IEEEbib}
\bibliography{IC2E}




\end{document}